\documentclass[aps,pra,twocolumn,groupedaddress]{revtex4-1}

\usepackage{amssymb}
\usepackage{amsthm}
\usepackage{amsmath}
\usepackage{graphicx}

\begin{document}

\title{Low and intermediate energy stopping power of protons and antiprotons in canonical targets}
\author{C. C. Montanari}
\email{mclaudia@iafe.uba.ar}
\author{J. E. Miraglia}
\email{miraglia@iafe.uba.ar}
\affiliation{Consejo Nacional de Investigaciones Cient\'{\i}ficas y
T\'{e}cnicas - Universidad de Buenos Aires, Instituto de
Astronom\'{\i}a y F\'{\i}sica del Espacio, Pabell\'on IAFE, 1428 Buenos Aires, Argentina;}%
\affiliation{Universidad de Buenos Aires, Facultad de Ciencias
Exactas y Naturales, Departamento de F\'{\i}sica, Ciudad
Universitaria, 1428 Buenos Aires, Argentina.}

\date{\today}

\begin{abstract}
In this work we propose a non-perturbative approximation to the
electronic stopping power based on the central screened potential of
a projectile moving in a free electron gas, by Nagy and Apagyi
[Phys. Rev. A \textbf{58} (1998) R1653]. We used this model to
evaluate the energy loss of protons and antiprotons in ten solid
targets: Cr, C, Ni, Be, Ti, Si, Al, Ge, Pb, Li and Rb. They were
chosen as \textit{canonicals} because they have reliable
Wigner-Seitz radius, $r_s$=1.48 to 5.31, which cover most of the
possible metallic solids. Present low velocity results agree well
with the experimental data for both proton and antiproton impact.
Our formalism describes the binary collision of the projectile and
one electron of the free electron gas. It does not include the
collective or plasmon excitations, which are important in the
intermediate and high velocity regime. The distinguishing feature of
this contribution is that by using the present model for low to
intermediate energies and the Lindhard dielectric formalism for
intermediate to high energies, we describe the stopping due to the
free electron gas in an extensive energy range. Moreover, by adding
the inner-shell contribution using the shellwise local plasma
approximation, we were able to describe all the available
experimental data in the low, intermediate and high energy regions.
\end{abstract}


\maketitle

\section{Introduction}

The energy loss of ions in solids has historically been a subject of
interest due to its importance in different fields of technological
and biological interest, such as ion beam analysis, radiation
damage, and range of ions in matter. The relevance of this subject
can be noted in the extended compilation of experimental data in
\cite{Paul}, very active up to the present time.

The theoretical developments cover from the Bethe theory for high
energies in the 30s \cite{Bethe} up to the time-dependent DFT for
very low impact velocities \cite{DFT,Ech07,Kohanoff12,Shukri16},
going through the dielectric formalism by Lindhard
\cite{Lindhard,LW,LS} and later models
\cite{Abril98,MM1,Montanari02}, the free electron gas models
\cite{ferrell77,Eche90}, and the binary theories
\cite{Sigmund,Grande,Arista02,Bailey15a}. Very effective too are the
semiempirical descriptions and codes, the most extended of which is
SRIM \cite{SRIM}. Many reviews on this subject have been published,
see for example the classic ones by Fano \cite{Fano63} and Inokuti
\cite{Inokuti71}, or more recently by Arista \cite{Arista04} and
Sigmund \cite{Sigmund17}.

In the last decade, the stopping power has had a revival due to the
requirement of more accurate experimental data, and to the
possibilities and precision of the up to date techniques
\cite{SDB_HCI16}. Perhaps  the most  challenging ones are the
low-energy antiproton experiments at CERN and the future prospects
of the Facility for Antiproton and Ion Research \cite{Widmann,FAIR}
at Darmstadt.

Based on the publications of experimental works compiled in
\cite{Paul}, the number of measured ion-target systems has increased
from 74 in the period 2005-2008, to 96 in 2009-2012 and 158 in
2013-2016. The studied targets are approximately two-third compounds
(mainly oxides and polymers) and one-third atomic targets, with
special interest in the very low velocity range (i.e. $v<1$). This
revival is related not only to the direct interest in the stopping
powers but also to the inclusion of these values in simulations with
different purposes \cite{Geant4,Barradas}. It must be mentioned that
most of the values included in these simulations come from SRIM
\cite{SRIM}, or the ICRU reports \cite{icru}, and important
discrepancies have been reported
\cite{Barradas,SDB_HCI16,Wittmaack}.

The impulse of the new experimental measurements of stopping by low
energy projectiles (i.e. by antiprotons
\cite{Moller3,Moller1,Moller2} and by protons \cite{
Vs94,HB06,Fm02,Roth13,Pr11b}) beards the theoretical developments.
The expected linear dependence with the velocity, the influence of
d-electron excitation and the density of electrons involved in the
projectile loss of energy, have attracted many of the stopping power
experimental efforts in the last years
\cite{Fig07,Ca09,Mar09,Go13,Go14,Cel15,Bauer17}. The theoretical
work on low energy stopping is extensive,  such as by the groups of
Echenique \cite{Eche81,Zaremba95}, Nagy \cite{Nagy98,Nagy96,Nagy94},
Arista \cite{Arista02,Arista98,FV_Arista,FV_Arista2},
Cabrera-Trujillo \cite{Cab00}, and more recently Kadyrov
\cite{Bailey15a,Bailey15b} and Grande \cite{Grande2016}.

The accuracy of the new experimental techniques and the necessity of
full theoretical data, lead us to wonder which is the highest
theoretical precision to describe these low-energy new experimental
measurements. For this purpose, we present here a non-perturbative
binary collisional model to describe the electronic stopping power
$dS/dx$ of heavy charged projectiles in a free electron gas (FEG).
The description at low impact velocities $v$ is amplified by
calculating the friction parameter $Q=(dS/dx)/v$. In order to have a
clear view of the problem to solve, we analyze the case of protons
and antiprotons (no charge state considerations), and targets of
well-established Wigner-Seitz radius, $r_s$.

We define \textit{the canonical} metallic solids as those of
reliable $r_S$, thus any doubts arising from their description can
be dispelled. The criterion we followed is that the theoretical
$r_{S}$ obtained considering the atomic density and the number of
valence electrons do not defer more than $5\%$ with respect to the
value deduced from the experimental plasmon energy \cite{Isaacson}.
In this way, we state these \textit{canonical} targets as settings
for future theoretical and experimental comparisons.

We present in this work a non-perturbative binary collisional model
to describe the electronic stopping power of heavy charged
projectiles in a free electron gas (FEG). The present model is based
on the central screened potential of a projectile moving in a free
electron gas, by Nagy and Apagyi \cite{Nagy98}, corrected in order
to verify the cusp condition. Thus, we have successfully faced the
theoretical problem of negative induced density by negative charge
intruders \cite{Singwi68}.

The main characteristics of our proposal are:
i) the use of a central potential $V_{Z}(r)$ that is Coulombic at
the origin and decays exponentially at large distances;
ii) an induced density that verifies the closure relation, which is
finite at the origin and never becomes negative (as it happens if we
use the Yukawa potential);
iii) the cusp condition is imposed through an additional parameter
$\lambda$.
This strategy is valid at low energies, or at least where plasmons
play a minor role. It only accounts for the outer electrons, so that
inner-shell contributions have to be included in an independent
form.

The goal of this work is to describe the stopping power of ions in
solids in an extended energy region. For this purpose, we resort to
two different descriptions for the valence electron contribution:
the present binary non-perturbative model for low and intermediate
energies, and the perturbative dielectric formalism in the
intermediate to high energy region. The dielectric formalism is
perturbative but contains not only binary but also the collective
excitations. The inner-shell contribution is included by using the
proved shellwise local plasma approximation (SLPA)
\cite{MM1,Cantero09,Pb}.

We chose ten canonical targets, Cr, C, Be, Ti, Si, Al, Ge, Pb, Li
and Rb, of Wigner-Seitz radius $r_{S}=1.48-5.31$, covering most of
the metallic solids. These targets belong to the groups of alcaline
metals (Li, Rb, Be), the post transition metals (Al, C, Si, Ge, Pb),
and the first groups of the transition metals (Ti and Cr). Among the
transition metals, those elements with few d-electrons also have
well-established $r_s$ values (groups 3 to 6 of the periodic table
of elements). Instead, we skip here groups 7 to 12, where
$d$-electrons play a quasi FEG role depending on the impact
velocity. There are very interesting targets that have been an
object of extensive experimental research in the last decade,
focused on targets such as Pd, Pt, Cu, Ag, Cu, Au, or Zn
\cite{Fig07,Ca09,Mar09,Go13,Go14,Cel15,Bauer17}.

In this contribution we only consider targets for which there are
low energy experimental data in the literature. For proton impact we
use the compilation in \cite{Paul}, for antiproton impact the
measurements by M\"oller and collaborators
\cite{Moller3,Moller1,Moller2}.

We describe the present formalism in section \ref{s2}. In section
\ref{s3} we show the scope of the model by comparing it with the low
energy measurements for protons and antiprotons. We extend the
theoretical-experimental comparison from 0.25 to 500 keV by
combining the binary and the dielectric formalisms for the FEG, and
the SLPA for the inner shells. The experimental needs and future
prospects are discussed in section \ref{4.}. Atomic units are used
in all this paper, except when it is explicitly stated.

\section{Theory}

\label{s2}

\subsection{Potential and density}
\label{2.A}

Consider a heavy bare Coulomb projectile of charge $Z$\ and velocity
$v$ travelling within a FEG. Let us model the projectile-electron
interaction by means of the central effective potential $V_{Z}(r)$
introduced by Nagy and Apagyi \cite{Nagy98}:
\begin{equation}
V_{Z}(r)=-\frac{Z}{r}\left( V_{1}e^{-\mu _{1}r}+V_{2}e^{-\mu
_{2}r}\right) ,  \label{10}
\end{equation}%
with
\begin{equation*}
\left\{
\begin{array}{cc}
V_{1} & =\frac{(\alpha +\beta )^{2}}{4\alpha \beta },\ \ \ \ \ \ \mu
_{1}=\alpha -\beta , \\
&  \\
V_{2} & =-\frac{(\alpha -\beta )^{2}}{4\alpha \beta },\ \ \ \ \mu
_{2}=\alpha +\beta%
\end{array}%
\right. ,
\end{equation*}%
This screened potential tends exponentially to zero at large
distances and has the correct limit $V_{Z}(r)\rightarrow -Z/r$ as
$r\rightarrow 0$ for any value of $\alpha $ and $\beta $.

The induced density $n_{i}$ can be determined by using the Poisson
equation to get
\begin{equation}
n_{i}(r) =Z \ \frac{(\alpha ^{2}-\beta ^{2})^{2}}{16\pi \alpha \beta \ r}%
(e^{-\mu _{1}r}-e^{-\mu _{2}r}).  \label{40}
\end{equation}%
It can be easily checked that $n_{i}$ verifies the desired closure
relation
\begin{equation}
\int d\overrightarrow{r}\ n_{i}(r)=Z,  \label{50}
\end{equation}%
as far as $Re(\alpha +\beta )>0$, and that it is finite at $r=0$,
\begin{equation}
n_{i}(r) =\frac{Z}{8\pi }(\alpha ^{2}-\beta ^{2})^{2}\left(
\frac{1}{\alpha }-r\ \right) +O(r^{2}).  \label{42}
\end{equation}%

\begin{figure}
\resizebox{0.53 \textwidth}{!} {
\includegraphics{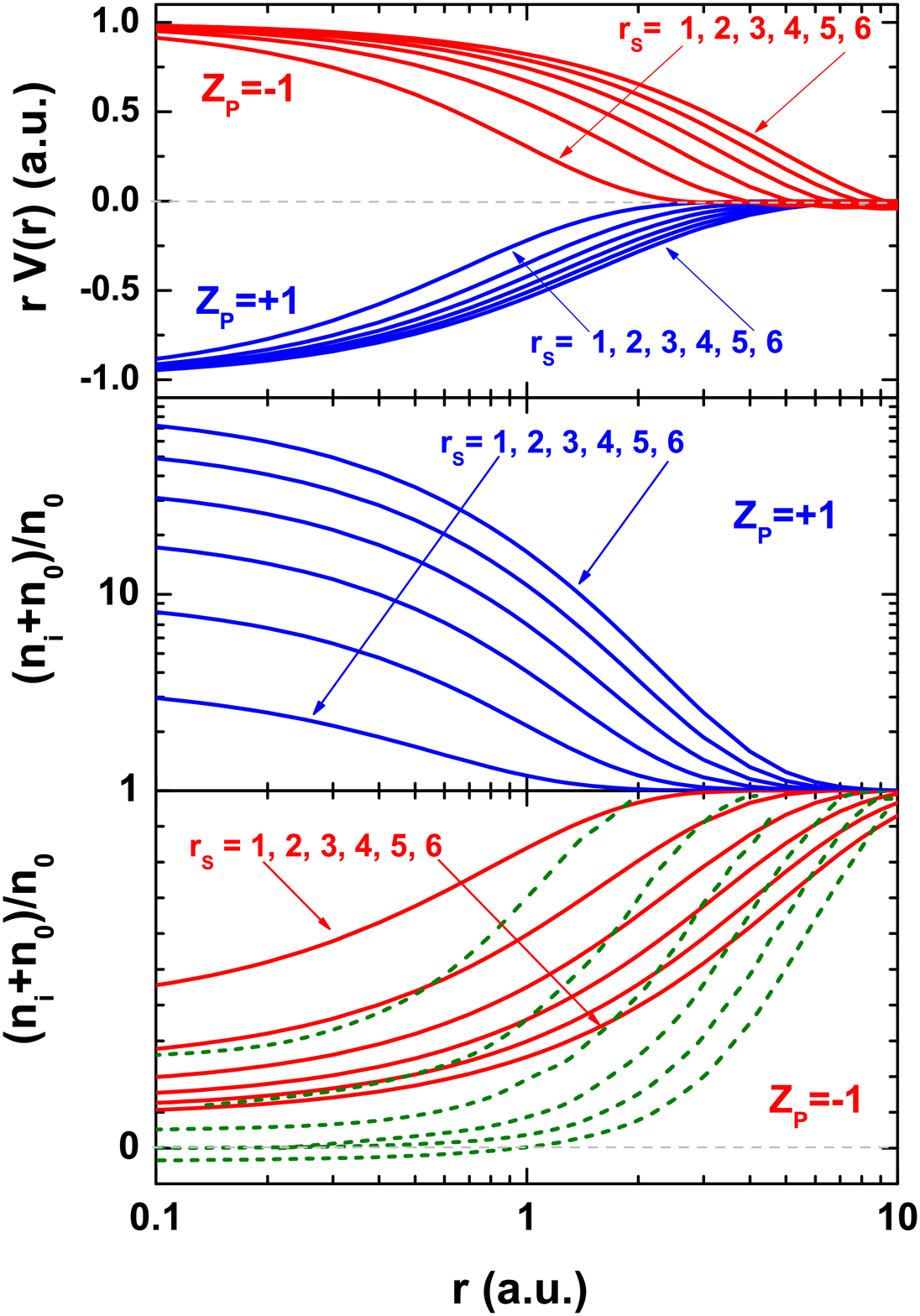}}
\caption{Screening potentials and electronic densities generated by
protons ($Z_P=+1$) and antiprotons ($Z_P=-1$) at rest in a FEG. We
display it for different values of $r_S=1-6$ as indicated inside the
figure. Curves: Solid-lines, present results, which verify the cusp
condition at the origin, and have positive densities for any value
of $r_S$; dashed-lines, the results for antiprotons in a FEG by
Singwi \protect\cite{Singwi68}. \label{fig1}}
\end{figure}

Following Nagy and Echenique \cite{Nagy93}, the parameters $\alpha $
and $\beta $ are defined as
\begin{eqnarray}
\alpha &=&\sqrt{b/\lambda +\omega _{P}/\sqrt{\lambda }},  \label{60} \\
\beta &=&\sqrt{b/\lambda -\omega _{P}/\sqrt{\lambda }},  \label{70}
\end{eqnarray}%
where $b$ can be related to the real part of the Lindhard dielectric
function,
\begin{equation}
b=\frac{v_{F}^{2}/3}{\frac{1}{2}+\frac{v_{F}^{2}-v^{2}}{4v\
v_{F}}\log \left\vert \frac{v+v_{F}}{v-v_{F}}\right\vert },
\label{90}
\end{equation}%
with $v_F$ being the Fermi velocity, $v_F=1.917\ r_S^{-1}$, and
$\omega _{P}=1.732\ r_S^{-3/2}$ is the plasmon frequency. As in
\cite{Nagy93}, we leave $\lambda $ as a free parameter determined by
imposing the cusp condition to the density,
\begin{equation}
-2Z=\underset{r\rightarrow 0}{\lim }\
\frac{\frac{d}{dr}n_{i}(r)}{n_{i}(r)+n_{0}}.  \label{80}
\end{equation}%
Note that Eq. (\ref{90}) introduces the dependency of the potential
with the ion velocity. The asymptotic limits of (\ref{90}) are
\begin{equation}
b\rightarrow \left\{
\begin{array}{cc}
v_{F}^{2}/3, & \ \ \ as\ \ v\rightarrow 0 \\
\\
v^{2}\ \ \ , & \ \ \ as\ \ v\rightarrow \infty
\end{array}%
\right. \label{95}
\end{equation}%

The cusp condition greatly improves  the behavior of $n_{i}$ at the
origin, erasing nonphysical negative electronic densities. For
example, in the case of antiprotons at rest in a FEG, $\lambda =1$
gives negative density, $n_{i}(0)+n_{0}<\ 0$, for certain values of
$r_{S}$. Instead, by imposing the cusp condition (\ref{80}) to get
$\lambda$, very reasonable values are obtained, i.e.
$-n_{0}<n_{i}(0)<n_{0}$. It is worth mentioning that, no matter the
impact velocity, the electronic density always verifies the cusp
condition at the origin.

In Fig. \ref{fig1} we plotted the potentials and total electronic
densities (solid lines) generated by protons ($Z=+1$) and
antiprotons ($Z=-1$) at rest. We consider $r_{S}$ ranging from $1$
to $6$, which covers by far all the known values. As expected, the
potential tends to be Coulombic as $r_{S}$ increases (the density of
electrons decreases, and therefore the screening too). The induced
densities displayed in Fig. \ref{fig1} satisfy the cusp condition
imposed by $\lambda $. As can be noted, even the densities
originated by antiprotons never become negative. The shape of the
density induced by a negative charge can also be interpreted as a
pair distribution function. For the sake of comparison, we include
in Fig. \ref{fig1} the pair distribution function reported by Singwi
\textit{et al.} \cite {Singwi68}. It is the best Random Phase
Approximation (RPA), including short-range correlation and exchange
effect. Even so, this RPA pair distribution function presents
negative densities at the origin for $r_{S}>4$.

At high velocities ($v>>v_{F}$), $\lambda \rightarrow 1$, and the
following expected limits are verified:
\begin{equation}
V(r)\underset{v\rightarrow \infty }{\rightarrow }-\frac{Z}{r}\
e^{-\frac{ \omega _{P}}{v}\ r},  \label{91}
\end{equation}
and
\begin{equation}
n_{i}(r=0)+n_{0}\underset{v\rightarrow \infty }{\rightarrow }\
n_{0}\ \left(1+ 2\ \frac{Z}{v}\right).  \label{101}
\end{equation}
Beyond the theoretical validity of these limits, the physics
involved in the present model only describes binary collisional
processes. The collective electronic excitations, also known as
plasmon excitations, are not included. The plasmon contribution is
important at high energies. Within the dielectric formalism, the
minimum impact velocity $v_P$ to excite plasmons can be approximated
as \cite{MM00}
\begin{equation}
v_{P}/v_{F} \simeq 1+(3\pi v_{F})^{-1/2}. \label{vc}
\end{equation}
We will return to this point in section \ref{s3} where we compare
our results (using the present non-perturbative model and using the
Linhard dielectric formalism) to the existing experimental data.

\subsection{Stopping power and friction}

The calculation of stopping power, or energy loss per unit path
length, $dS/dx$, implies the integration of the electron momentum
$\overrightarrow{k}_{i}$ over all the Fermi sphere \cite{ferrell77},
\begin{equation}
\frac{dS}{dx}(v)=2\int \frac{d\overrightarrow{k}_{i}}{\left( 2\pi \right)
^{3}}\ \theta (k_{i}-k_{F})\ \frac{ds}{dx}(\overrightarrow{k^{\prime }}_{i})
\label{200}
\end{equation}%
with
\begin{equation}
\frac{ds}{dx}(\overrightarrow{k^{\prime }}_{i})=2\pi \frac{k_{i}^{\prime }}{v%
}\ \overrightarrow{k_{i}^{\prime }}\cdot \overrightarrow{v}\ \sigma _{tr}(%
\overrightarrow{k_{i}^{\prime }})  \label{210}
\end{equation}%
and $\overrightarrow{k_{i}^{\prime }}=\overrightarrow{k_{i}}-\overrightarrow{%
v}$ the relative velocity. The transport cross section $\sigma
_{tr}(k)$ is
\begin{equation}
\sigma _{tr}(k)=\frac{4\pi }{k^{2}}\sum\limits_{l=0}^{\infty }(l+1)\sin ^{2}%
\left[ \delta _{l}(k)-\delta _{l+1}(k)\right]   \label{220}
\end{equation}%
with $\delta _{l}(k)$ being the phase shifts generated by the
potential $V_{Z}(r)$. The central potential given by (\ref{10}) is
expressed in terms of exponentials, so the first Born approximation
to $\sigma _{tr}(k)$ can be calculated straightforwardly.

An alternative expression to the transport cross section has been
recently proposed  by Grande \cite{Grande2016}, derived from the
retarding force due to an asymmetrically-induced charge density
acting on the projectile. This non-central density is calculated
from the spherically symmetric potential using the partial-wave
expansion in a frame fixed to the ion. The results by Grande
\cite{Grande2016} are interesting, mainly in the intermediate energy
region, which is the conflictive zone where the validity of the high
energy models (including plasmons) and the low energy ones (binary)
compete.

The stopping power can also be expressed in terms of the friction
parameter $Q_Z(v)$ as
\begin{equation}
\frac{dS}{dx}=Z^{2} \ v \ Q_Z(v) , \label{230}
\end{equation}%
According to Fermi and Teller, at low impact velocities the stopping
power is expected to show a linear dependency with the velocity, and
so $Q_Z(v)$ tends to a constant. In the perturbative regime the
stopping power is proportional to $Z^2$, so $Q_Z(v)$ is independent
of $Z$.

The linear response theory (LRT) by Ferrel and Ritchie
\cite{ferrell77}, predicts
\begin{equation}
Q^{LRT}(v \to 0)=\frac{2}{3\pi }\left( \ln \left(
1+\frac{6.03}{r_{S}}\right) - \frac{1}{1+\frac{r_{S}}{6.03}}\right).
\label{240}
\end{equation}
This is a first perturbative approximation, therefore insensitive to
the charge $Z$ of the intruder.

Taking into account the projectile charge and velocity and the
screening by the FEG, a reasonable criterion for the perturbative
regime is
\begin{equation}
v/v_F \ \geq \ Z_P \ r_S.
\label{244}
\end{equation}
In fact, as $v_F=1.917/r_S$, then this criteria is equivalent to $v
\geq 1.917 \ Z_P$. We will return to this in the next section in
view of the theoretical-experimental comparison.

\section{Results and comparison with the experimental data}
\label{s3}

In this section we display the results of the present formalism for
antiproton and proton impact in Cr, C, Be, Ti, Si, Al, Ge, Pb, Li,
and Rb. We chose them because they are typical canonical metals,
i.e. their valence electrons act as a free electron gas of constant
and well known value of $r_S$. As we will comment later in this
work, there are many more metallic targets that could be described
using the present model, but there are no experimental measurements
at low energies to compare with (see section \ref{4.}).

In table I we list the ten targets, their $r_S$ and $v_F$ values
\cite{Isaacson}, the calculated $v_P$ \cite{MM00}, and our
non-perturbative results for $Q_{+1}(v)$ and $Q_{-1}(v)$ in the
limit $v \to 0$. These values may be used as predictions for future
low energy measurements by proton and antiproton impact.

\begin{table}
\caption{The 10 solid targets studied here, their Wigner-Seitz
radii, $r_S$, and Fermi velocity, $v_F$ \cite{Isaacson}, the
calculated minimum velocity for plasmon excitation, $v_{P}$, given
by (\ref{vc}), and the present results for the friction in the limit
$v\to 0$ for proton, $Q_{+1}(0)$, and antiproton $Q_{-1}(0)$ impact,
as defined in (\ref{230}). Atomic units are used throughout this
table. \label{table1}}

\begin{ruledtabular}
\begin{tabular}{|c|c|c|c|c|c|c|}
& $r_{S}$ & $r_{S}^{exp}$ & $v_{F}$ & $v_P/v_F$ & $Q_{+1}(0)$ &
$Q_{-1}(0)$
\\ \hline
$Cr$ & 1.48 & 1.55 & 1.30  & 1.29 & 0.307 & 0.176 \\
$C$  & 1.60 & 1.66 & 1.20  & 1.30 & 0.295 & 0.163 \\
$Be$ & 1.87 & 1.78 & 1.03  & 1.32 & 0.269 & 0.141 \\
$Ti$ & 1.92 & 1.93 & 1.00  & 1.33 & 0.264 & 0.137 \\
$Si$ & 2.01 & 1.97 & 0.955 & 1.33 & 0.256 & 0.131 \\
$Al$ & 2.07 & 2.12 & 0.927 & 1.34 & 0.250 & 0.127 \\
$Ge$ & 2.09 & 2.02 & 0.918 & 1.34 & 0.248 & 0.126 \\
$Pb$ & 2.30 & 2.26 & 0.834 & 1.36 & 0.229 & 0.113 \\
$Li$ & 3.27 & 3.21 & 0.587 & 1.43 & 0.150 & 0.075 \\
$Rb$ & 5.31 & 5.45 & 0.361 & 1.54 & 0.041 & 0.048 \\ 
\end{tabular}
\end{ruledtabular}
\end{table}

\subsection{Proton and antiproton energy loss at low impact velocities}
\label{3.1}

\begin{figure}
\resizebox{0.53\textwidth}{!} {
\includegraphics{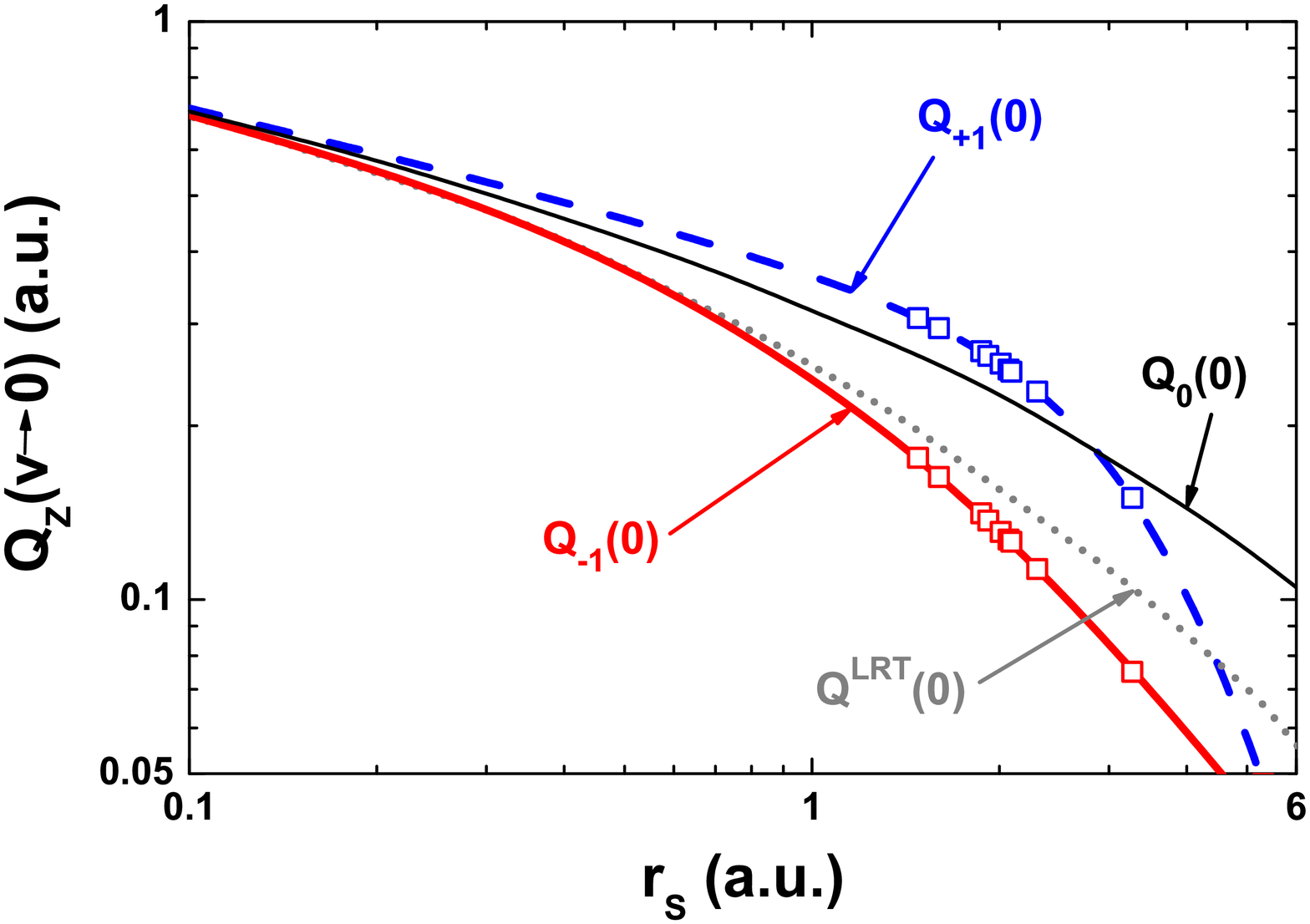}
} \caption{The friction in the limit $v \to 0$,  as a function of
$r_S$. Curves: present results for antiprotons, $Q_{-1}$, solid-red
line; protons, $Q_{+1}$ dashed-blue line; and $Q_0$ given by Eq.
(\ref{Z0}), thin black line; and grey dotted-line, $Q^{LRT}(v)$ in
the linear response theory by Ferrel and Ritchie \cite{ferrell77}
given by (\protect\ref{240}). Symbols: hollow squares are the
theoretical values displayed in table \ref{table1} for specific
targets. \label{fig3}}
\end{figure}

In what follows we compare our friction $Q_Z(v)$ at low impact
energies with the experimental data available in the literature. We
focus on this energy region in order to have only the contribution
of the valence electrons. In Fig. \ref{fig3} we report our results
for proton $Q_{+1}$, and for antiproton impact $Q_{-1}$, in the
limit $v \to 0$ as a function of $r_{S}$. The theoretical values
displayed in table \ref{table1} for specific targets are also
included in Fig. \ref{fig3}. These results confirm the experimental
evidence that protons cede more energy to the FEG than antiprotons.

We also include in Fig. \ref{fig3} (dotted line) the prediction of
the linear response theory (LRT) by Ferrel and Ritchie
\cite{ferrell77}, Eq. (\ref{240}), which is independent of the
projectile charge. The comparison of our non-perturbative values and
the linear ones is very interesting. Our results correctly tend to
the $Q_{LRT}$ as $r_{s}\rightarrow 0$, where we can consider that
the screening of the projectile is so high that it can be described
as a perturbation. On the contrary, as $r_{S}$ increases, the
projectile becomes a huge perturbation to the FEG so the linear
models cannot describe it.

To explore the perturbative limit we also calculated the friction
for $Z \to 0$,
\begin{equation}
Q_0(v)=\underset{Z\to 0}{Lim}\ Q_Z(v) \label{Z0}
\end{equation}
and plotted it in Fig. \ref{fig3} (black solid line). For $r_S>1$,
$Q_0$ divides the region between $Q_{+1}$ and $Q_{-1}$, the known
Barkas effect \cite{Sigmundv1}.

\begin{figure}[tbp]
\resizebox{0.53 \textwidth}{!} {
\includegraphics{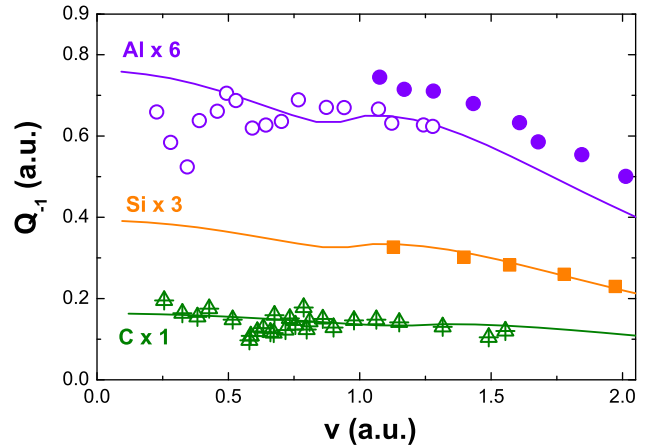}
} \caption{The friction parameter $Q_Z$ as a function of the impact
velocity for antiprotons in C, Si and Al. Curves, present
non-perturbative results; symbols, experimental data: for Al, empty
circles \cite{Moller3}, solid circles \cite{Moller2}; for Si, solid
squares \cite{Moller2}; and for C, crossed triangles
\cite{Moller1}.} \label{fig4}
\end{figure}

The description of the energy loss by antiproton impact is a
challenge for any model. But it has the advantage that there is no
possibility of charge exchange \cite{Widmann}. The measurements by
Moeller \textit{et al.} \cite{Moller3,Moller1,Moller2} for
antiprotons in several targets let us to test our theory with the
Coulomb sign of the intruder. In Fig. \ref{fig4} we display the
present values for the friction as a function of the impact velocity
for antiprotons in C, Si, and Al. Note that the agreement is very
good in a linear-scale plot. We focused on the low velocity region
in order to have only the valence electron contribution. Inner
shells, however, may be contributing for Al above $v=1.5$. The
theoretical description of low-energy antiproton measurements is an
open path for future experimental research \cite{Widmann}. Energy
loss investigation is part of the physics program of the
next-generation antiproton source FLAIR (Facility for Low-energy
Antiproton and Ion Research) \cite{FLAIR}, planned for the next five
years.

\begin{figure}[tbp]
\resizebox{0.53\textwidth}{!} {
\includegraphics{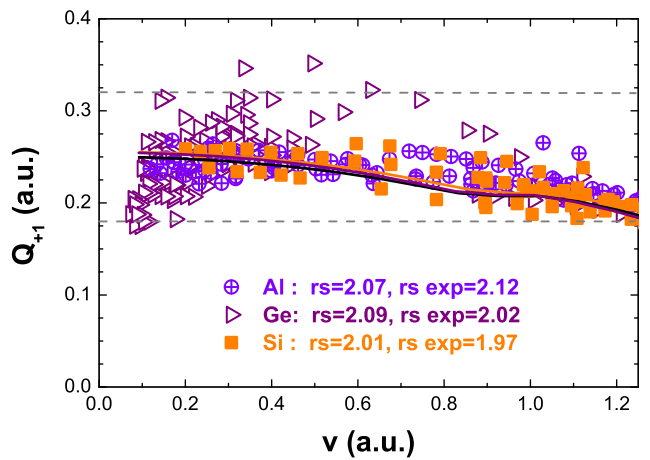}
} \caption{The friction parameter $Q$ as a function of the impact
velocity for protons in Ge, Al and Si. Curves, present
non-perturbative results; symbols, experimental data available, as
compiled in \protect\cite{Paul}.} \label{figrs2}
\end{figure}

At low impact energies the stopping power depends only on the value
of $r_S$, so it should be the same for different metals of similar
$r_S$. In Fig. \ref{figrs2} we plot together the experimental data
for protons in three targets of $r_{S} \simeq 2$, Al, Ge and Si. As
can be noted in this figure, all the low energy measurements are
quite close within the experimental spread. We also display in this
figure our theoretical results for Al, Si and Ge, which are actually
very close and nicely describe the low energy measurements in the
three targets. This confirms that $r_S$ is the \textit{only}
relevant parameter at low energies.

The experimental frictions displayed in Fig. \ref{figrs2} are
$Q_{+1}^{exp}=0.25\pm 0.07$, and so are our theoretical results,
with $Q_{+1} \to 0.25$ as $v\to 0$. Similar values of the friction
at low energies are expected for other metals of $r_{S}\simeq 2$,
such as Zn, Ga or Te. The recent low energy measurements for protons
in Zn \cite{Go14} confirm this. But the most interesting point is
the prediction of the value of $Q_{+1}$ for future measurements in
Te, with no data of stopping at all, or in Ga, with no low energy
measurements \cite{Paul}.

\subsection{Low to intermediate energy region}

\begin{figure}
\resizebox{0.53\textwidth}{!} {
\includegraphics{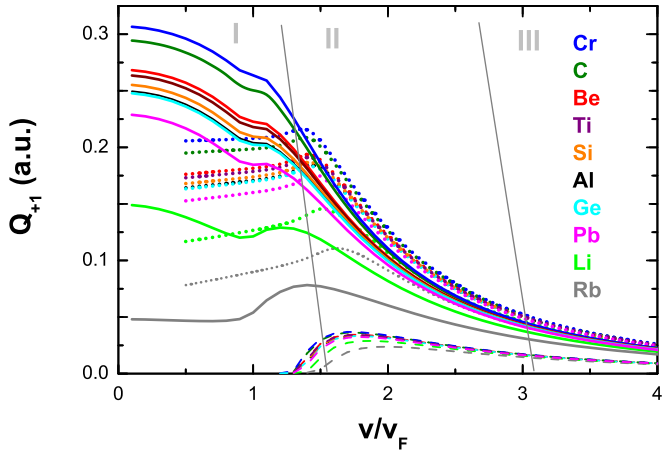}
} \caption{[color online] Friction for protons in different targets
as a function of the ratio of the impact velocity and the Fermi
velocity. Curves: solid-lines, the present non-perturbative results;
dotted-lines, the Lindhard dielectric formalism results;
dashed-lines, the isolated plasmon excitation contribution included
in the calculations using the dielectric formalism. The different
targets are plotted following the order: from top to bottom they are
Cr, C, Be, Ti, Al, Si, Ge, Li, and Rb.} \label{fig2p}
\end{figure}

Above certain impact velocity $v_{P}$, the energy loss implies not
only binary but also collective excitations (plasmons)
\cite{ferrell77}. Though the present non-linear binary theory has
the correct high energy limit, in the intermediate energy region it
lacks the collective contribution. The extension to impact
velocities $v>v_{P}$ can be performed by using the well-known
Lindhard dielectric formalism \cite{Lindhard,LW}. This formalism
includes both binary and collective excitations, and tends to the
Bethe limit at high energies. But it is a linear response
approximation, therefore valid within the perturbative limits.

A detailed comparison of the friction as a function of the impact
velocity by using the present model (non-perturbative) and the
Lindhard dielectric formalism (perturbative) is presented in Fig.
\ref {fig2p}. The results for protons in Cr, C, Be, Ti, Si, Al, Ge,
Pb, Li and Rb, are displayed in this figure from top to bottom. Note
that the lowest the $r_S$ the largest the stopping power. Three
regions I, II and III, are indicated in Fig. \ref {fig2p}, separated
at $v_{P}$ and $2\ v_{P}$. In region I ($v<v_{P}$), the present
collisional formalism (solid lines) is very appropriate because it
includes all the perturbative orders and only binary collisions are
involved. Instead, the dielectric formalism (dotted lines) is very
poor, it underestimates the stopping power in this region. On the
contrary, in region III ($v>2\ v_{P}$), which is clearly
perturbative, the dielectric formalism is correct since it includes
both plasmons and single-electron excitations. The non-perturbative
results are below the dielectric ones in this region. Region II is
the intermediate one, in which non-linear effects and plasmon
contribution compete in importance. The validity of each formalism
in this region is subject to the comparison with the experimental
data, as will be shown in section \ref{3.2}.

We also include in Fig. \ref {fig2p} the isolated plasmon
contribution (dashed lines). Note that the impact velocity $v_P$
above which the plasmon excitation starts contributing agrees quite
well with the values in table I. It can be observed that for
$v>v_{P}$ plasmon excitation is not negligible at all. As predicted
by Lindhard and Winter \cite{LW}, at very high energies the
equipartition rule holds and the binary stopping equals to the
plasmon one.

\subsection{Extension to higher energies, the inner-shell contribution}
\label{3.C}

\begin{figure}
\resizebox{0.53\textwidth}{!} {
\includegraphics{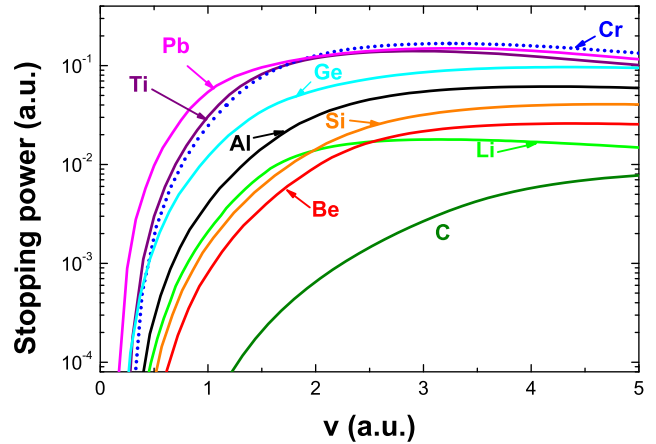}
} \caption{[color online] Inner-shell contribution to the stopping
power of protons in Cr, Pb, Ti, Ge, Al, Si, Li, Be and C, as
function of the impact velocity. Curves, results obtained with the
SLPA considering from the K-shell to the sub-valence one, according
to each target. \protect\cite{MM1,Cantero09,Pb}.} \label{fig2}
\end{figure}

At sufficiently high impact energies the impinging projectile will
be able to remove sub-valence electrons. To extend the theoretical
description to intermediate and high velocities we include the
inner-shell contribution by resorting to the SLPA
\cite{MM1,Cantero09,Pb}. During the last years we have developed
this model based on the dielectric formalism and the local plasma
approximation by Lindhard and Scharff \cite{LS}. The contribution of
each sub-shell of target electrons is described including screening,
collective response and correlation in the final state. The inputs
are the densities and binding energies of each sub-shell. For
non-relativistic atoms, i.e. atomic numbers up to 54, they can be
obtained from the Hartree-Fock wave functions by Bunge \textit{et
al.} \cite{Bunge}. For targets of higher atomic numbers, the
relativistic Dirac equation must be solved. The most interesting
characteristic of the SLPA is that it is a density-based model,
therefore capable of being used for molecular targets as far as a
good description of the different shells electronic density is
available \cite{TiO2,ZnO}. The great limitation is that it is a
perturbative model.

In Fig. \ref{fig2} we display the SLPA results for the stopping
power due to the inner-shells of Cr, C, Be, Ti, Al, Si, Ge, Li and
Pb. For Pb (Z=82, relativistic target) we used the results obtained
by employing the \rm{GRASP} code in \cite{Pb}. For the rest, we used
the atomic wave functions by Bunge \cite{Bunge}. As we are dealing
with solids, the binding energies are slightly different from those
of the gas phase. We use the experimental binding energies relative
to the top of the Fermi level for metals compiled by Williams \cite
{Will95}, instead of the theoretical values for single atoms, which
correspond to gases.

It can be noted in Fig. \ref{fig2} that the inner-shell contribution
falls down several orders of magnitude when going from high to low
energies. We are fully aware of the inability of the perturbative
SLPA to describe the low energy region, but inner-shell contribution
is relatively negligible in this energy region. On the contrary, as
velocity increases, the relative importance of the inner-shells
grows and, at the same time, the validity of the SLPA starts to
hold.

\subsection{Comparison with the experiments in an extended energy range}
\label{3.2}

We performed an extensive comparison of the present theoretical
results and the experimental data in the IAEA database \cite{Paul}.
We analyzed the stopping of protons in Cr, C, Be, Ti, Si, Al, Ge, Pb
and Li. We did not include Rb in this comparison because there are
no measurements in the low energy region, which is our main
interest. By combining our non-perturbative and perturbative
calculations in different energy regions, we managed to cover an
extended range of $(0.25-500)$ keV. The extension to higher impact
energies by using the dielectric formalism and the SLPA has already
been demonstrated \cite{MM1,Cantero09,Pb}.

\begin{figure}[tbp]
\resizebox{0.53\textwidth}{!} {
\includegraphics{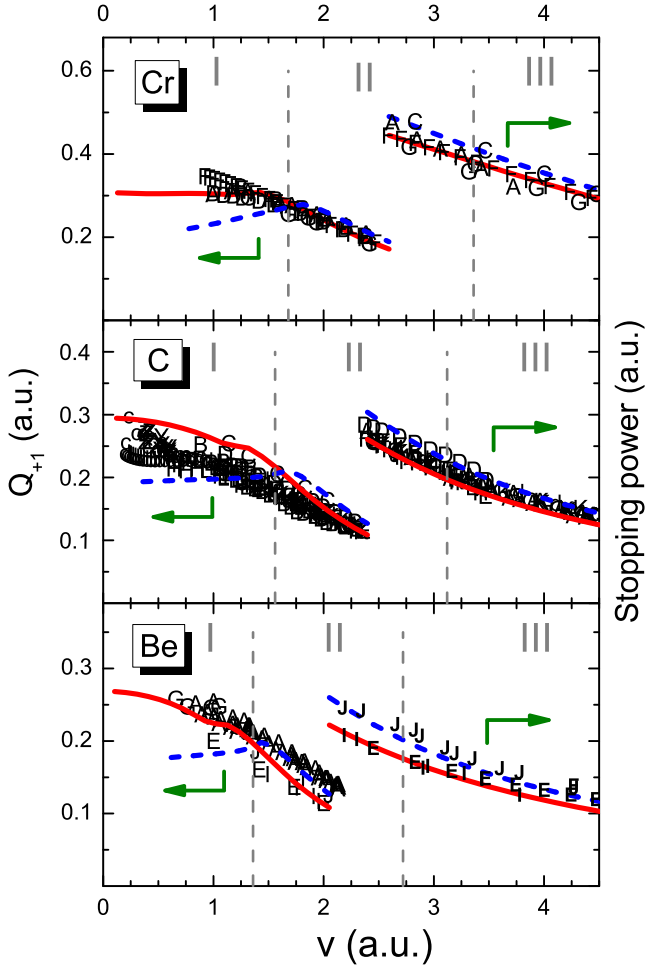}
} \caption{Total friction ($v<1.5 \ v_P$) and stopping power ($v>1.5
\ v_P$) (including FEG and inner shells) as function of the impact
velocity, for protons in Cr, C and Be. Curves: red solid lines,
present results using the non-perturbative model for the FEG; blue
dashed lines, present values using the Lindhard dielectric function
for the FEG (linear response). In both cases the inner-shell
contribution is included, calculated with the SLPA
\protect\cite{MM1}. Symbols: letters, available experimental data in
\protect\cite{Paul} and references therein. } \label{fig:C_Cr_Be}
\end{figure}

\begin{figure}[tbp]
\resizebox{0.53\textwidth}{!} {
\includegraphics{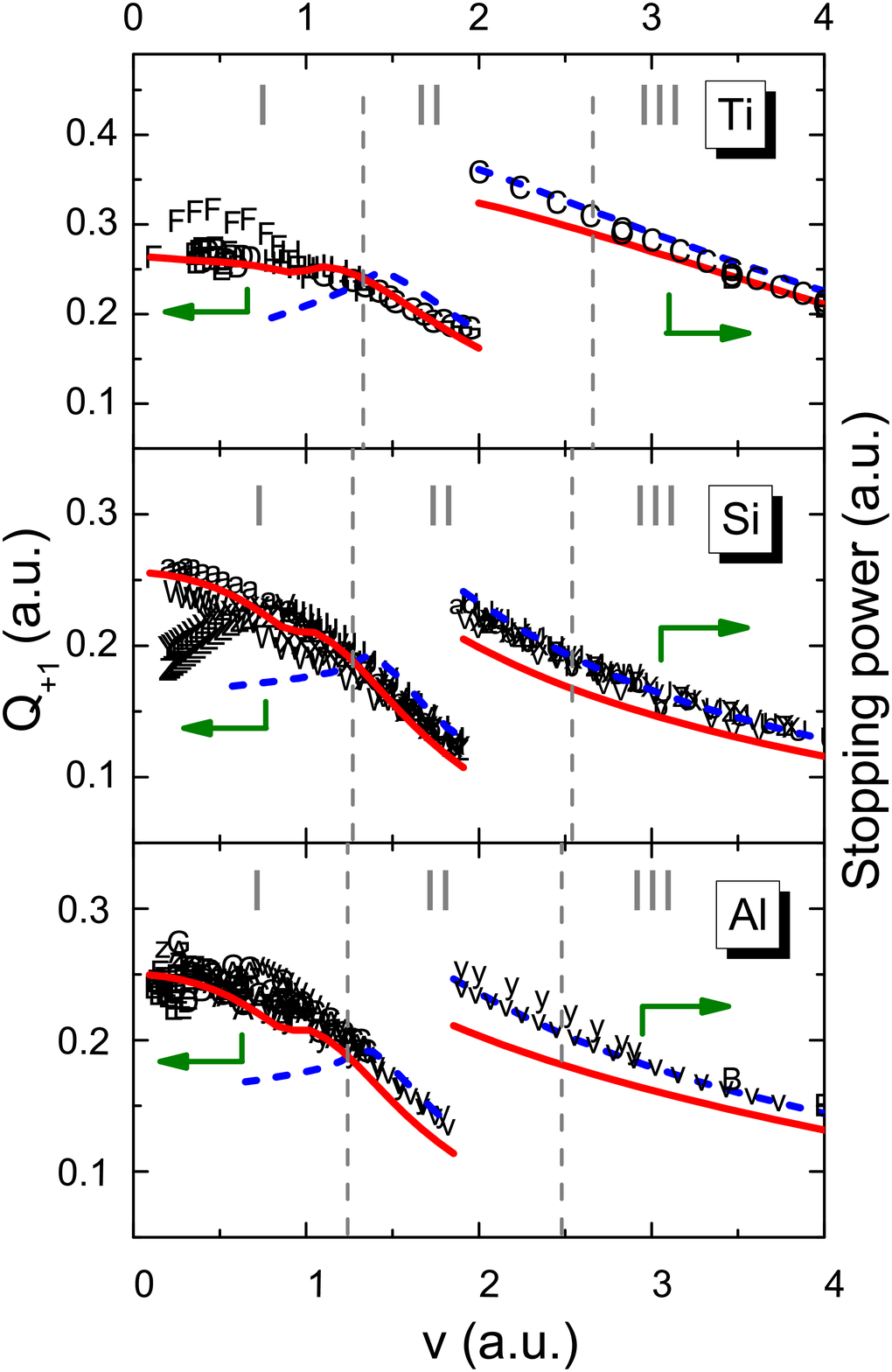}
} \caption{Total friction ($v<1.5 \ v_P$) and total stopping power
($v>1.5 \ v_P$) (including FEG and inner shells) as function of the
impact velocity, for protons in Ti, Si, and Al. Curves and symbols
as in \protect\ref{fig:C_Cr_Be}. For Si and Al the whole available
data in \cite{Paul} is abundant. We only include here the data since
1990. We added recent data for H in Al by Moeller \textit{et al.}
\cite{Moller3} (letter G), that was not in \cite{Paul}.}
\label{fig6bis}
\end{figure}

\begin{figure}[tbp]
\resizebox{0.53\textwidth}{!} {
\includegraphics{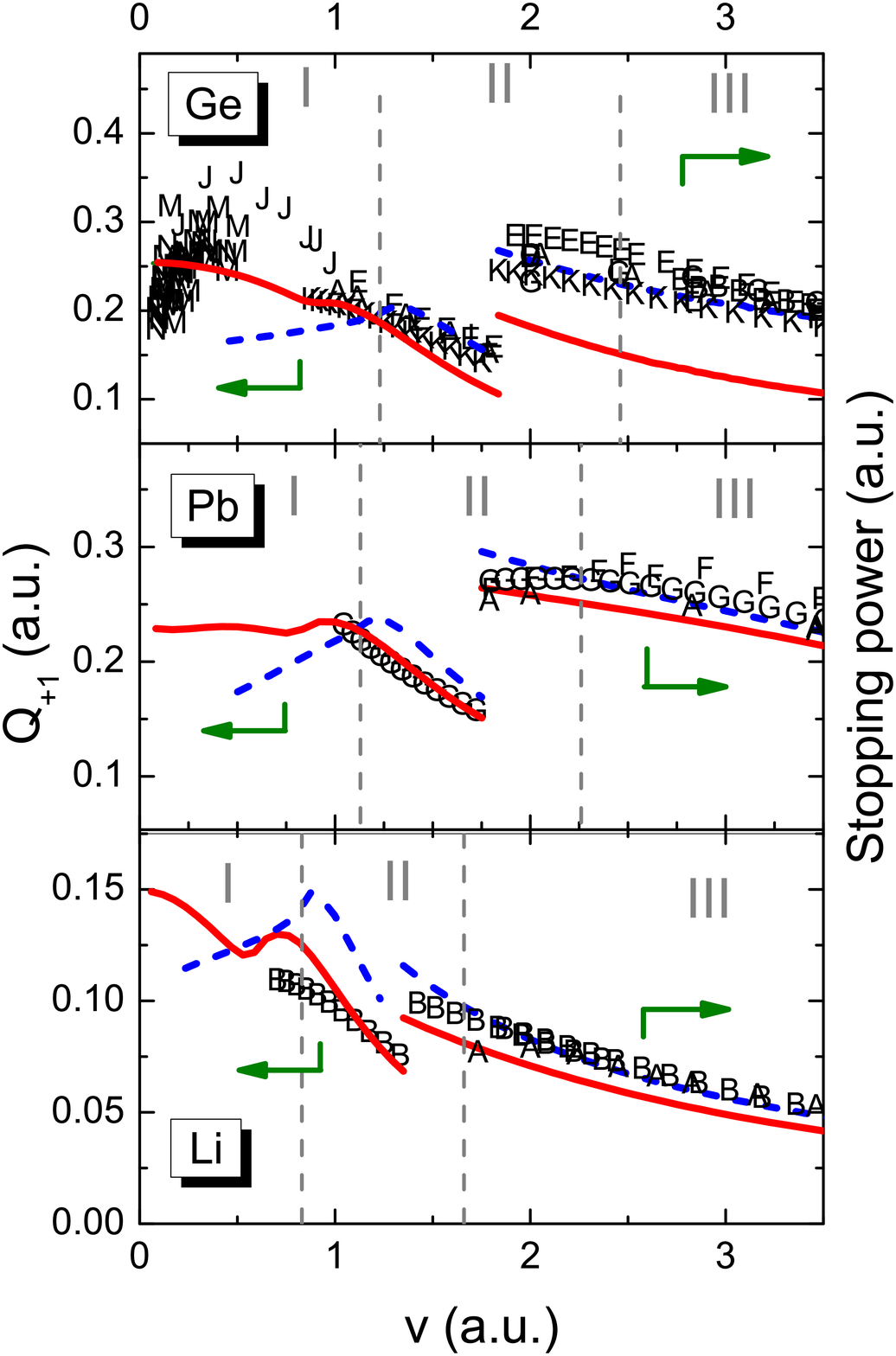}
} \caption{Total friction ($v<1.5 \ v_P$) and total stopping power
($v>1.5 \ v_P$) (including FEG and inner shells) as function of the
impact velocity, for protons in Ge, Pb and Li. Curves and symbols as
in \protect\ref{fig:C_Cr_Be}.} \label{fig7bis}
\end{figure}

In Figs. \ref{fig:C_Cr_Be}-\ref{fig7bis} we compare our theoretical
results with the available experimental data \cite{Paul} for the
nine targets mentioned above. We display the friction for the lowest
velocities in order to heighten the low stopping values. Instead,
for the highest velocities we plotted the stopping power.

We show in these three figures the total values using the
non-perturbative approximation for the FEG (red solid lines), and
the perturbative model for the FEG (blue dashed lines). The total
stopping power was obtained by adding the SLPA results for the
inner-shell contribution. The experimental data in Figs.
\ref{fig:C_Cr_Be}-\ref{fig7bis} follow the same notation using
different letters as symbols as in \cite{Paul}.

We separate three regions as in Fig. \ref{fig2p}. Both boundaries,
at $v_P$ and $2\ v_P$, are displayed with vertical dashed-lines.
These energy regions involve different physical regimes. In the low
energy region I, valence electrons are the main contribution and a
non-perturbative description is mandatory. The high energy region
III corresponds to the perturbative regime. As mentioned before, the
intermediate region II is very interesting because plasmon
excitation starts to occur and the validity of the perturbative
description will depend on each case. It is worth to note that the
stopping maximum is in this region, for impact velocity $v \gtrsim
v_{P}$.

Figure \ref{fig:C_Cr_Be} displays the present results for Cr, C and
Be ($r_S=1.48, 1.6$ and $1.87$, respectively).  In the upper plot,
for protons in Cr, our non-perturbative results clearly describe the
experimental measurements in the whole energy range. The data by
Eppacher and Semrad \cite{Ep92} of 1992 (represented by letter "F"
in this plot) is the most recent one and covers an extended energy
region from $20-700$ keV. Only the low energy values, i.e. impact
velocity below 1.2, seem to be too large. There is no experimental
data for $v<1$. New low energy measurements for this system are
welcome. On the other hand, the FEG of Cr has the highest electronic
density, or equivalently, the smallest $r_{S}$ considered here. This
implies a large screening of the projectile, and almost a
perturbative regime in the whole velocity range. This explains the
agreement of the perturbative calculations down to impact velocities
$v \geq \ 1.7$.

Also displayed in Fig. \ref{fig:C_Cr_Be} are the present results for
stopping of protons in amorphous Carbon. This is one of the targets
with more experimental measurements due to its different
applications. However, the complexity of carbon (amorphous or
cristal phases) also introduces dispersion among different sets of
measurements. We show in this figure the available data since 1980.
It can be noted that our non-perturbative friction reproduces the
experiments in regions II and III, but overestimates a little in
region I. As predicted, the perturbative results are reasonable for
$v \geq 1.917 \  Z_P$. Note that for carbon we also reproduce very
well the antiproton impact measurements, even for very low
velocities (see Fig. \ref{fig4}).

Finally, for Beryllium, at the bottom part of Fig.
\ref{fig:C_Cr_Be}, we begin to note the difference between the
non-perturbative and the perturbative descriptions, the latter
including plasmons. The separation between the curves is clear for
$v>v_{P}$. The agreement of the present non-perturbative results
with the low-energy data is very good. However for $v\geq 1.5$ the
non-perturbative results are too low, while the perturbative ones
describe well the experiments. This difference is explained by the
lack of plasmon contribution in the binary model. The experimental
values represented by letters "E" and "I" correspond to the
measurements by Warshaw \cite{warshaw} and by Kahn \cite{kahn} in
the 50s, which are below the general tendency of much recent data
\cite{Paul}. It can be said that the combination of the present
non-perturbative model for $v \leq v_P$ and the dielectric formalism
for $v \geq v_P$ gives a good description of the energy loss of
protons in Be in the whole energy range.

Figure \ref{fig6bis} displays the present results for Ti, Si and Al.
Again the vertical dashed lines separate the three energy regions
mentioned above. In the upper figure, the energy loss in Ti is
nicely described in the whole energy range, showing a very good
agreement with the experiments. For low velocities, $v<1.8$ a.u.,
the present non-perturbative formalism describes the data correctly.
Only the low energy measurements by Arkhipov \textit{et al.} in 1969
\cite{Ar69} (letter "F") are higher than the rest. This detail would
not be noticed if we plotted the stopping power instead of the
friction coefficient at low energies. Some doubts on the
normalization of Arkhipov's data have been stated by Paul in
\cite{Paul}. Titanium is a target of technological importance that
deserves new stopping measurements, not only in the low energy
region, but also around the stopping maximum, where only one set of
data is available (by Ormrod in 1971 \cite{Or71}, letter "G" in this
figure). In the intermediate region II, the binary model clearly
underestimates the experimental values for $v\geq 2$. This can be
adjudicated to the lack of plasmons, included in the dielectric
formalism.

The energy loss of protons in Si have more than 600 experimental
values for the different energies. Among all these data, we show in
Fig. \ref{fig6bis} those measured since 1990. It is a criteria to
have a clearer view of the experimental tendency and to avoid the
great dispersion among the oldest measurements. The agreement of our
results for protons in Si shown in Fig. \ref{fig6bis} is very good
from the very low to the high energies. This is more noticeable if
we focus on the latest experimental measurements: the low energy
data by Hobbler \textit{et al.} in 2006 \cite{HB06} (letter "a" in
region I), by Fama \textit{et al.} in 2002 \cite{Fm02} (letter "W"
in regions I and II), and the high energy data by Abdesselam
\textit{et al.} in 2008 \cite{Ab08} (letter "b" in regions II and
III). Instead, the measurements by Konac \textit{et al.} in 1998
\cite{Kc98} (letters "Y" and "Z") are too low for $v<0.7$. Again
this difference in the friction would be very small if we plot the
stopping power in the low velocity regime too. The friction plot
acts as a magnifier of the low velocity behavior, which is very
demanding for any theoretical description. We describe Si as a free
electron gas with no energy gap, and we do not extend the
calculations below $v=0.1$. We do not discuss here the threshold of
the Si as semiconductor \cite{Roth13,Archubi16}, which is below
$v=0.03$.  For $v>v_P$ the perturbative calculation improves the
binary one. The combination of formalisms, the non-perturbative one
in region I and the dielectric one in regions II and III, lead us to
correctly describe the experimental data of stopping power of
protons in Si in the whole energy range. This includes the stopping
maximum, which is around 52 keV ($v=1.5$). For Si, we obtain very
good agreement with the experiments, not only for protons (Fig.
\ref{fig6bis}) but also for antiprotons (Fig. \ref{fig4}).

In the bottom plot of Fig. \ref{fig6bis} we display the
theoretical-experimental comparison for Aluminum, which is one of
the most studied targets. As for the previous case, we restrain the
comparison to modern experiments (1990 up to now). The agreement at
low energies is quite nice, specially with the newest low energy
data by Primetzhofer \textit{et al.} \cite{Pr11b} in 2011 (with
letters "E" and "D" in Fig. \ref{fig6bis}). Clearly the perturbative
model underestimates the friction in this region. The results in
regions II and III show that for $v>v_P$ the non-perturbative
formalism (binary) underestimates the measurements, while the
dielectric results (binary and plasmos) nicely describe the data.
Our proposal is that the combination of both models allows to
describe the stopping of H in Al in the whole energy range, from the
very low up to the high energies.

Finally, in Fig. \ref{fig7bis} we plot the energy loss of protons in
Ge, Pb and Li. We can note in these cases that all the physics
involved is enhanced, i.e. the importance of a non-linear
description as compared to the contribution of plasmons. For the
case of Ge, in the upper part of Fig. \ref{fig7bis}, the validity
ranges of the non-perturbative binary formalism and the dielectric
formalism (binary+plamons) are very clear. The description of the
experimental data by the non-perturbative calculation in region I is
quite good. It nicely links the data by Eppacher and Semrad
\cite{Ep92}, (with letter "K"), and the most recent measurements
(letters "N" and "M") by Bauer and collaborators in Linz
\cite{Roth13}. On the other hand, the data by Arkhipov and Gott in
1969 \cite{Ar69} (letter "J") is clearly above the rest. Present
results also agree with the TDFDT values for protons in Ge
\cite{Ullah15}, which apply only to very low energies (i.e. $v\leq
0.6$). In regions II and III, our binary formalism underestimates
the measurements showing the importance of plasmon excitations in
these energy regions. Instead, the perturbative results describe
nicely the measurements for $v>v_P$, and clearly separate from the
binary results. This behavior is the expected one, it has been
already found for Si and Al in Fig. \ref{fig6bis}, but for Ge the
difference is more pronounced.

The medium graphic in Fig. \ref{fig7bis} corresponds to protons in
Pb. This case is special because we are dealing with a relativistic
target, 82 bound electrons, including the K-L-M-N-O shells and the
6s$^2$-6p$^2$ electrons as free electron gas. We follow \cite{Pb} to
calculate the inner-shell contribution to the energy loss by using
the SLPA together with the relativistic densities of electrons of
each subshell (spin-orbit split) obtained with the GrasP code. In
the present calculations, the experimental binding energies were
used \cite{Will95}. This improves our previous results in \cite{Pb}
in the high energy region.

The $r_{S}=2.3$ of Pb is higher than in the previous targets, hence
it is more non-perturbative.  This can be noted in the comparison of
both models (solid and dashed curves) in the intermediate region II.
We can say that the non-linear contributions are more important than
the plasmons for Pb. The binary non-perturbative calculations for Pb
clearly improve the perturbative ones for $v \leq \ 1.5 \ v_{P}$,
with very good agreement with Eppacher data \cite {Ep92}.
Unfortunately, there are no measurements of stopping of protons in
lead below 25 keV. Our results indicate an almost linear tendency of
the stopping of protons in Pb for $v \leq 0.7$, with $Q_{+1}\sim
0.23$. Low energy measurements would be a good test for this
prediction.

Finally, we display in the bottom part of Fig. \ref{fig7bis} our
results and the experimental data for protons in Li. This is the
target with the largest value of $r_S=3.27$ we here considered, and
so the smallest electron density and the lowest Fermi velocity,
$v_{F}=0.59$. This makes Li a very interesting test of our model
because it is highly non-perturbative around the stopping maximum,
i.e. between $20-40$ keV. Only two sets of data are available for
this system \cite{Paul}, so more measurements are welcome, mainly
for $v<1.5$. The present non-perturbative model describes properly
the experimental values in the intermediate region II, but it
overestimates the data for $v < 1$. It is worth to mention that
different theoretical calculations for H in solid Li by Kaneko
\cite{K93}, and by Cabrera Trujillo \textit{et al} \cite{Cab08} are
also above the measurements by Eppacher \textit{et al.} \cite{Ep95}
around the stopping maximum. In the high energy region III, the
perturbative model is valid, and the good description of the
experiments for impact velocities $v>1.7$ shows the importance of
plasmon excitations.

\section{Experimental scarcity and future prospects} \label{4.}

The great absent in the comparisons of section \ref{3.2} is
Rubidium, with a FEG characterized by $r_{S}=5.31$ and a very low
Fermi velocity, $v_F=0.361$.  Protons introduce a huge perturbation
to such a FEG, which can test any non-perturbative theoretical model
to the limit. Unfortunately, the available measurements are for
$v\geq 1.1$, which is 3 times $v_F$. This turns Rb a very
interesting target to be studied, experimentally and theoretically,
and an opportunity of future research.

We expect the predictions for low-velocity friction as a function of
the $r_S$ presented here (Fig. \ref{fig3} and table \ref{table1}) to
be benchmarks for future measurements. There are more canonical
metals \cite{Isaacson}. In general, these targets belong to the s
and p-blocks of the periodic table of elements (alkaline metals and
earth metals, with valence s-electrons; post transition metals and
metalloids, with valence p-electrons). However, for many of them
there are no experimental stopping powers at low impact energies.
For example, for proton impact in the s-block elements there is no
data for impact energies $E<20$ keV for Mg ($r_{S}$=2.66), Ca
($r_{S}$=3.27), and Sr ($r_{S}$=3.59), and there is no data at all
for Na ($r_{S}$=3.99). Also some relativistic targets, such as Cs
($r_{S}$=5.75) and Ba ($r_{S}$=3.74) have no stopping data at all.
Among the elements of the p-block of metals, there is no low energy
data for protons in Ga ($r_{S}$=2.19) for $E<70$ keV, in Sn
($r_{S}$=2.4) for $E<20$ keV, and there is no data at any impact
energy for protons in Se ($r_{S}$=1.84) and Te ($r_{S}$=2.09). Note
that the latter is a very interesting case to test the universal
predictions of Fig. \ref{figrs2} for the $r_S \simeq 2$ elements.

The transition metals of groups 7 to 12 of the periodic table have
been the focus of attention for the low energy experimental research
during the last fifteen years. Unexpected experimental changes were
found in friction when d-electrons start to be active in the
collisions. It can be thought as an inhomogeneus $r_S$, depending on
the impact velocity. However, even for the transition metals, those
elements of groups 3 to 6 (the d sub-shells mostly empty) have
canonical $r_s$ values and could be tested by our non-perturbative
model if low energy stopping data were available. Some examples are
V ($r_{S}$=1.66), with no data for $E<30$ keV, Nb ($r_{S}$=3.07)
with no data for $E<20$ keV and great dispersion of the experimental
data around the maximum of the sopping power, and Mo ($r_{S}$=1.61)
with no data for $E<70$ keV. Even the relativistic W ($r_{S}$=1.62)
has no data for $E<80$ keV. An interesting case is Ta, with very
recent measurements for $E<10$ keV by Bauer \textit{et al}
\cite{Bauer17}, and an unexplained high density of valence
electrons. This target requires the relativistic treatment to
determine the shell to shell electronic densities and binding
energies for the SLPA calculation of the stopping by inner shells.

All the targets mentioned above are interesting aims for future
experimental and theoretical research. Knowing their stopping values
is important, not only as atomic solids, but also because they are
known partners in compounds of technological interest
\cite{SDB_HCI16}, and most of the stopping calculations in compounds
are obtained from their components, with bond corrections in some
cases. So reliable predictions of their values would be very useful.

\section{Conclusions}

In this work we propose a non-linear model to deal with low and
intermediate impact stopping based on a central screened potential
for a projectile moving in a free electron gas. This potential
induces a density of electrons that verifies the cusp condition at
the origin, independently of the impact velocity, and the charge
sign of the intruder.

In order to test this model for proton and antiproton impact we
chose canonical solid targets (reliable value of the Wigner-Seitz
radius $r_S$), with experimental data available in the low energy
region: Cr, C, Be, Ti, Si, Al, Ge, Pb, and Li. The comparison at low
impact velocities was done in terms of the friction (stopping power
per impact velocity), which is a very sensitive parameter, and let
us to test the linear dependency with the velocity.

We proved that the present non-perturbative model gives a good
description of the low energy data for antiprotons in C, Si and Al,
and for protons in Cr, Be, Ti, Si, Al, Ge and Pb. For protons in C
and Li some small overestimation is found as discussed in the text.

By combining the present model for low to intermediate energies, and
the dielectric formalism (including plasmons) for intermediate to
high energies, a good description of the stopping power was obtained
in an extended energy range. The inner shell contribution was
included by using the perturbative SLPA model. A detailed
theoretical-experimental comparison was performed considering all
the data available. We analyzed our results in three energy regions:
for low impact energies up to that of plasmon excitations (the
non-perturbative regime); for high energies (the perturbative
regime), and in the intermediate energy region. We showed that in
this intermediate region the non-perturbative description and the
plasmon excitation compete in importance, depending on the $r_S$. We
suggested that the perturbative description is valid for $v/v_F \geq
r_S\ Z_P$. However, we found that for  $r_S < 2.1$ the perturbative
results are valid even for lower impact velocities, $v/v_F\simeq 1.3
\ Z_P$.

We recall the importance of Rb as a highly non-perturbative case
(very low $v_F$) with no low energy measurements. We have also
detected at least thirteen elements of well-known $r_S$  but with
unmeasured stopping power at low energies. These targets deserve
future experimental and theoretical research.

\begin{acknowledgments}
This work was supported by the following institutions of Argentina:
Consejo Nacional de Investigaciones Cient\'{\i}ficas y T\'ecnicas,
Agencia Nacional de Promoci\' on Cient\'{\i}fica y Tecnol\' ogica,
and Universidad de Buenos Aires. The authors acknowledge Pedro
Grande for useful comments on this work.

\end{acknowledgments}



\begin{thebibliography}{99}

\bibitem{Paul}
\textit{Stopping Power of Matter for Ions, Graphs, Data, Comments and Programs},
https://www-nds.iaea.org/stopping/.

\bibitem{Bethe} H. Bethe,
\textit{Zur Theorie des durchgangs schneller Korpuskularstrahlen
durch materie} , Ann. Phys. \textbf{5}, 325 (1930).

\bibitem{DFT} M. A. L. Marques and E. K. U. Gross,
\textit{Time-dependent density functional
theory}, Annual Review of Physical Chemistry (2004), DOI:
10.1146/annurev.physchem.55.091602.094449

\bibitem{Ech07} M. Quijada, A. G. Borisov, I. Nagy, R. Díez Muiño,
and P. M. Echenique,
\textit{Time-dependent density-functional
calculation of the stopping power for protons and antiprotons in
metals}, Phys. Rev. A \textbf{75}, 042902 (2007).

\bibitem{Shukri16} A.A. Shukri, F. Bruneval, and L. Reining,
\textit{Ab initio electronic stopping power of protons in bulk
materials}, Phys. Rev. B \textbf{93},  035128 (2007).

\bibitem{Kohanoff12} M. Ahsan Zeb, J. Kohanoff, D. S\'anchez-Portal, A. Arnau, J.I. Juaristi, and Emilio
Artacho,
\textit{Electronic Stopping Power in Gold: The Role of d
Electrons and the H=He Anomaly}, Phys. Rev. Lett. \textbf{108},
225504 (2012).

\bibitem{Lindhard} J. Lindhard,
\textit{On the properties of a gas of charged particles},
Mat. Fys. Medd. Dan. Vid. Selsk \textbf{28}, 1-57  (1954).

\bibitem{LW} J. Lindhard and A. Winter,
\textit{Stopping power of elelctron gas and equipartition rule},
Mat. Fys. Medd. Dan. Vid. Selsk \textbf{34}, 1-22  (1964).

\bibitem{LS} J. Lindhard and M. Scharff,
\textit{Energy Dissipation by Ions in the kev Region},
Phys. Rev. \textbf{124}, 128 (1961).

\bibitem{Abril98} I. Abril, R. Garcia-Molina, C. D. Denton, F. J. P\'erez-P\'erez, and
N. R. Arista,
\textit{Dielectric description of wakes and stopping
powers in solids}, Phys. Rev. A \textbf{58},  357 (1998).

\bibitem{MM1} C. C. Montanari and J. E. Miraglia,
\textit{The Dielectric Formalism for
Inelastic Processes in High-Energy Ion–Matter Collisions}, Adv.
Quant. Chem. \textbf{65}, edited by Dz. Belkic (Elsevier, Amsterdam,
2013), Chap. 7, pp. 165-201.

\bibitem{Montanari02} C. C. Montanari, J. E. Miraglia, andN.R. Arista,
\textit{Dynamics of solid inner-shell electrons in collisions with
bare and dressed swift ions}, Phys. Rev. A \textbf{66}, 042902
(2002).

\bibitem{ferrell77} T. L. Ferrell and R. H. Ritchie,
\textit{Energy losses by slow ions and atoms to electronic
excitation in solids}, Phys. Rev. B \textbf{16}, 115 (1977).

\bibitem{Eche90} P. M. Echenique, F. Flores, and R. H. Ritchie,
\textit{Dynamic screening of ions in matter}, Solid State Phys.
\textbf{43} 229 (1990).

\bibitem{Sigmund} P. Sigmund and A. Schinner,
\textit{Binary stopping theory for swift heavy ions}, Eur. Phys. J.
D 12,  425 (2000).

\bibitem{Grande} G. Schiwietz and P. L. Grande, Nucl. Instrum. Methods Phys. Res.  B
\textit{A unitary convolution approximation for the impact-parameter
dependent electronic energy loss} \textbf{153}, 1 (1999);
\textit{The unitary convolution approximation for heavy ions}
\textbf{195}, 55 (2002).

\bibitem{Arista02} N. R. Arista,
\textit{Energy loss of ions in solids:
Non-linear calculations for slow and swift ions}, Nucl. Instrum. Methods Phys. Res.  B \textbf{195}, 91 (2002).

\bibitem{Bailey15a} J. J. Bailey, A. S. Kadyrov, I. B. Abdurakhmanov, D. V. Fursa, and I.
Bray, \textit{Antiproton stopping in atomic targets}, Phys. Rev. A
\textbf{92}, 022707 (2015).

\bibitem{SRIM} J. F. Ziegler, M. D. Ziegler, J. P. Biersack,
\textit{SRIM – The stopping and range of ions in matter}, Nucl.
Instrum. Methods Phys. Res.  B \textbf{268}, 1818 (2010); SRIM code,
http://www. srim.org/.

\bibitem{Fano63} U. Fano,
\textit{Penetration of Protons, Alpha Particles, and Mesons}, Annual
Rev. Nucl. Science \textbf{13},1-66 (1963).

\bibitem{Inokuti71} M. Inokuti,
\textit{Inelastic collisions of fas charged particles with atoms and
molecules - The bethe theory revisited}, Rev. Mod. Phys. \textbf{43}
297-347 (1971).

\bibitem{Arista04} N. R. Arista and A. F. Lifschitz,
\textit{Non-Linear Approach to the Energy Loss
of Ions in Solids}, Adv. Quantum Chem. \textbf{45}, 47  (2004).

\bibitem{Sigmund17} P. Sigmund,
\textit{Six Decades of Atomic Collisions in Solids}, Nucl. Instrum.
Methods Phys. Res.  B (2017), in press,
http://dx.doi.org/10.1016/j.nimb.2016.12.004.

\bibitem{SDB_HCI16} C. C. Montanari and P. Dimitrou,
\textit{The IAEA stopping power database, following the trends in
stopping power of ions in matter}, Nucl. Instrum. Methods Phys. Res.
B (2017), in press, http://doi.org/10.1016/j.nimb.2017.03.138

\bibitem{Widmann} E. Widmann, \textit{Plans for a next-generation low-energy antiproton
facility}, Physica Scripta \textbf{72}, C51–C56 (2005).

\bibitem{FAIR}Facility for Antiproton and Ion Research,
http://www.fair-center.eu/.

\bibitem{Geant4} J. Allison \textit{et al.},
\textit{Recent developments in Geant4},
Nucl. Instrum. Meth. in Phys. Res. A \textbf{835}, 186 (2016), and
http://geant4.web.cern.ch/geant4/.

\bibitem{Barradas} N. P Barradas and E. Rauhala, \textit{Data analysis software for ion beam
analysis}, Joint ICTP/IAEA Workshop on Advanced Simulation and
Modelling for Ion Beam Analysis (2009).

\bibitem{icru} International Commission on Radiation Units and
Measurements, Rep. \textbf{37} (1984); Rep. \textbf{49 }(1993); Rep.
\textbf{73} (2005).

\bibitem{Wittmaack} K. Wittmaack,
\textit{On the origin of apparent Z1-oscillations in low-energy heavy-ion ranges},
Nucl. Instrum. Methods Phys. Res.  B \textbf{388}, 15 (2016).

\bibitem{Moller3} S.P. Moeller, A. Csete, T. Ichioka, H. Knudsen,
U.I. Uggerhoej, and H. H. Andersen, \textit{Stopping Power in
Insulators and Metals without Charge Exchange}, Phys. Rev. Lett
\textbf{93}, 042502 (2004).

\bibitem{Moller1} S.P. Moeller, A. Csete, T. Ichioka, H. Knudsen,
U.I. Uggerhoej, and H. H. Andersen, \textit{Antiproton Stopping at
Low Energies: Confirmation of Velocity-Proportional Stopping Power},
Phys. Rev. Lett \textbf{88}, 193201 (2002).

\bibitem{Moller2} S.P. Moeller, U.I. Uggerhoej, H. Bluhme,
H. Knudsen, U. Mikkelsen, K. Paludan, E. Morenzoni, \textit{Direct
measurements of the stopping power for antiprotons of light and
heavy targets}, Phys. Rev. A \textbf{56},  2930 (1997).

\bibitem{Vs94} J. E. Valdes, J. C. Eckardt, G. H. Lantschner, and N. R.
Arista, \textit{Energy loss of slow protons in solids: Deviation
from the proportionality with projectile velocity}, Phys. Rev. A
\textbf{49}, 1083 (1994).

\bibitem{HB06} G. Hobler, K. K. Bourdelle, T. Akatsu,
\textit{Random and channeling stopping power of H in Si below 100
keV}, Nucl. Instrum. Methods Phys. Res.  B \textbf{242}, 617 (2006).

\bibitem{Fm02} M. Fama, G. H. Lantschner, J. C. Eckardt, N. R. Arista, J. E.
Gayone, E. Sanchez, F. Lovey, \textit{Energy loss and angular
dispersion of 2–200 keV protons in amorphous silicon}, Nucl.
Instrum. Methods Phys. Res.  B \textbf{193}, 91 (2002).

\bibitem{Roth13} D. Roth, D. Goebl, D. Primetzhofer, P. Bauer,
\textit{A procedure to determine electronic energy loss from
relative measurements with TOF-LEIS},  Nucl. Instrum. Methods Phys.
Res.  B \textbf{317}, 61 (2013).

\bibitem{Pr11b} D. Primetzhofer, S. Rund,D. Roth, D. Goebl, P.
Bauer, \textit{Electronic Excitations of Slow Ions in a Free
Electron Gas Metal: Evidence for Charge Exchange Effects}, Phys.
Rev. Lett. \textbf{107}, 163201 (2011).

\bibitem{Fig07} E.A. Figueroa, E.D. Cantero, J.C. Eckardt, G.H. Lantschner,
J.E. Vald\'es, and N.R. Arista, %
\textit{Threshold effect in the energy loss of slow protons and
deuterons channeled in Au crystals}, Phys. Rev. A \textbf{75},
010901 (2007).

\bibitem{Ca09} E. D. Cantero, G.H. Lantschner, J.C. Eckardt, and N.R.
Arista, \textit{Velocity dependence of the energy loss of very slow
proton and deuteron beams in Cu and Ag}, Phys. Rev. A \textbf{80},
032904 (2009).

\bibitem{Mar09} S. N. Markin, D. Primetzhofer, M. Spitz, and P. Bauer,
\textit{Electronic stopping of low-energy H and He in Cu and Au
investigated by time-of-flight low-energy ion scattering}, Phys.
Rev. B \textbf{80}, 205105 (2009).

\bibitem{Go13} D. Goebl, K. Khalal-Kouache, D. Roth, E. Steinbauer,
and P. Bauer, \textit{Energy loss of low-energy ions in transmission
and backscattering experiments}, Phys. Rev. A \textbf{88}, 032901
(2013).

\bibitem{Go14} D. Goebl, W. Roessler, D. Roth, and P. Bauer,
\textit{Influence of the excitation threshold of d electrons on
electronic stopping of slow light ions}, Phys. Rev. A \textbf{90},
042706 (2014).

\bibitem{Cel15} C.E. Celedon \textit{et al.},
E. A. Sanchez, L. Salazar Alarcon, J. Guimpel, A. Cortes, P. Vargas,
and N. R. Arista, \textit{Band structure effects in the energy loss
of low-energy protons and deuterons in thin films of Pt}, Nucl.
Instrum. Methods Phys. Res.  B \textbf{360}, 103 (2015).

\bibitem{Bauer17} D. Roth, B. Bruckner, M. V. Moro, S. Gruber, D. Goebl,
J. I. Juaristi, M. Alducin, R. Steinberger, J. Duchoslav, D.
Primetzhofer, and P. Bauer,
\textit{Electronic Stopping of Slow
Protons in Transition and Rare Earth Metals: Breakdown of the Free
Electron Gas Concept}, Phys. Rev. Lett \textbf{118}, 103401 (2017).

\bibitem{Eche81} P. M. Echenique, R. M. Nieminen, and R. H. Ritchie, \textit{
Density functional calculation of stopping power of an electron gas
for slow ions}, Solid State Commun. \textbf{37} 779 (1981).

\bibitem{Zaremba95} E. Zaremba, A. Arnau, P.M. Echenique,
\textit{Nonlinear screening and stopping powers at finite projectile velocities}
Nucl. Instrum. Methods Phys. Res.  B \textbf{96}, 619  (1995).

\bibitem{Nagy98} I. Nagy and B. Apagyi, \textit{Scattering-theory
formulation of stopping powers of a solid target for protons and
antiprotons with velocity-dependent screening}, Phys. Rev. A \textbf{58}, R1653 (1998).

\bibitem{Nagy96} I. Nagy and A. Bergara,
\textit{A model for the velocity-dependent screening}, Nucl.
Instrum. Methods Phys. Res.  B \textbf{115}, 58 (1996).

\bibitem{Nagy94} I. Nagy, \textit{Low-velocity antiproton stopping,
A trial-potential approach},
Nucl. Instrum. Methods Phys. Res.  B \textbf{94}, 377 (1994).

\bibitem{Arista98} A. F. Lifschitz and N. R. Arista,
\textit{Electronic energy loss of helium ions in aluminum using the
extended-sum-rule method}, Phys. Rev. A \textbf{58}, 2168 (1998).

\bibitem{FV_Arista} J. M. Fern\' andez-Varea, and N.R. Arista,
\textit{ Analyticalformulaforthestoppingpoweroflow-energyions in
afree-electrongas}, Radiat. Phys. Chem. \textbf{96}, 88 (2014).

\bibitem{FV_Arista2} H. B. Nersisyan, J. M. Fern\' andez-Varea, and N. R.
Arista, \textit{Dynamic screening of an ion in a degenerate electron
gas within the second-order Born approximation}, Nucl. Instrum.
Methods Phys. Res.  B \textbf{354}, 167 (2015).

\bibitem{Cab00} R. Cabrera-Trujillo, Y. \"Ohrn, E. Deumens, and J. R.
Sabin, \textit{Stopping cross section in the low- to
intermediate-energy range: Study of proton and hydrogen atom
collisions with atomic N, O, and F}, Phys. Rev. A \textbf{62},
052714  (2000).

\bibitem{Bailey15b} J. J. Bailey, A. S. Kadyrov, I. B. Abdurakhmanov, D. V. Fursa, and I.
Bray, \textit{Antiproton stopping in H2 and H2O}, Phys. Rev. A
\textbf{92}, 052711 (2015).

\bibitem{Grande2016} P. L. Grande,
\textit{Alternative treatment for the energy-transfer and transport cross section
in dressed electron-ion binary collisions}, Phys. Rev. A \textbf{94}, 042704  (2016).

\bibitem{Isaacson} D. Isaacson, \textit{Compilation of rs values}, NewYork University
Rep. No. 02698 (National Auxiliary Publication Service, NY 1975).

\bibitem{Singwi68} K. S. Singwi, M. O. Tosi, R. H. Land, and A.
Sj\"olander, \textit{Electron Correlations at Metallic Densities},
Phys. Rev. \textbf{176}, 589  (1968).

\bibitem{Cantero09} E. D. Cantero, R. C. Fadanelli, C. C. Montanari, M. Behar,
J. C. Eckardt, G. H. Lantschner, J. E. Miraglia, and N. R. Arista
\textit{Experimental and theoretical study of the energy loss of Be
and B ions in Zn}, Phys. Rev. A \textbf{79}, 042904 (2009).

\bibitem{Pb} C. C. Montanari, C. D. Archubi, D. M. Mitnik, J. E. Miraglia,
\textit{Energy loss of protons in Au, Pb, and Bi using relativistic
wave functions}, Phys. Rev. A \textbf{79} 032903 (2009).

\bibitem{Nagy93} I. Nagy and P. M. Echenique,
\textit{Stopping power of an electron gas for antiprotons at
intermediate velocities}, Phys. Rev. A \textbf{47},  3050 (1993).

\bibitem{MM00} C. C. Montanari, J. E. Miraglia, and N. R. Arista,
\textit{Suppression of projectile-electron excitations in collisions
with a free-electron gas of metals}, Phys. Rev. A \textbf{62},
052902 (2000).

\bibitem{FLAIR}
http://www.fair-center.eu/public/experiment-program/appa-physics/flair.html;
http://www.flairatfair.eu/

\bibitem{Sigmundv1} P. Sigmund,  \textit{Particle Penetration and Radiation Effects,
General Aspects and Stopping of Swift Point Charges} vol. 1
(Springer-Verlag, Berlin, Heidelberg, 2006).

\bibitem{Bunge} C. F. Bunge, J. A. Barrientos, A. V. Bunge, and J. A.
Cogordan, \textit{Hartree-Fock and Roothaan-Hartree-Fock energies
for the ground states of He through Xe}, Phys. Rev. A \textbf{46},
3691 (1992).

\bibitem{TiO2} S.P. Limandri \textit{et al.},
\textit{Stopping cross sections of TiO2 for H and He ions}, Eur.
Phys. J. D \textbf{68}, 194 (2014).

\bibitem{ZnO} R.C. Fadanelli, C.D. Nascimento, C.C. Montanari, J.C. Aguiar, D. Mitnik,
A. Turos, E. Guziewicz, and M. Behar, \textit{Stopping and
straggling of H and He in ZnO}, Eur. Phys. J. D \textbf{70}, 178
(2016).

\bibitem{Will95} G. P. Williams,
\textit{Electron Binding Energies of the Elements},
CRC Handbook of Chemistry and Physics, Vol. F170 (CRC Press, Boca
Raton, 1986);
http://www.jlab.org/gwyn/ebindene.html.

\bibitem{Ep92} Ch. Eppacher,D. Semrad,
\textit{Dependence of proton and helium energy loss in solids upon plasma properties},
Nucl. Instrum. Methods Phys. Res.  B \textbf{69}, 33 (1992).

\bibitem{warshaw} S. D. Warshaw, \textit{The Stopping Power for Protons in Several Metals},
Phys. Rev. \textbf{76}, 1759 (1949).

\bibitem{kahn} D. Kahn, \textit{The Energy Loss of Protons in Metallic Foils and Mica},
Phys. Rev. \textbf{90}, 503 (1953).

\bibitem{Ar69} E. P. Arkhipov and Yu. V. Gott,
\textit{Slowing down of 0.5-30 keV protons in some materials},
Sov. Phys. -JETP \textbf{29}, 615 (1969).

\bibitem{Or71} J. H. Ormrod,
\textit{Electronic stopping cross sections of deutrons in titanium},
Nucl. Instrum. Methods  \textbf{95}, 49 (1971).

\bibitem{Ab08} M. Abdesselam, S. Ouichaoui, M. Azzouz, A. C. Chami, and
M.Siad, \textit{Stopping of 0.3–1.2 MeV/u protons and alpha
particles in Si}, Nucl. Instrum. Methods Phys. Res.  B \textbf{266},
3899 (2008).

\bibitem{Kc98} G. Konac, S. Kalbitzer, Ch. Klatt, D. Niemann, R.
Stoll, \textit{Energy loss and straggling of H and He ions of keV
energies in Si and C}, Nucl. Instrum. Methods Phys. Res.  B
\textbf{136-138} 159 (1998).

\bibitem{Archubi16} C.D. Archubi and N.R. Arista,
\textit{A study of threshold effects in the energy loss of slow
protons in semiconductors and insulators using dielectric and
non-linear approaches}, Eur. Phys. J. B \textbf{ 89}, 86 (2016).

\bibitem{Ullah15} R. Ullah, F. Corsetti, D. S\'anchez-Portal and E.
Artacho, \textit{Electronic stopping power in a narrow band gap
semiconductor from first principles}, Phys. Rev. B \textbf{91},
125203 (2015).

\bibitem{K93} T. Kaneko,
\textit{Partial and Total Electronic Stopping Cross Sections of Atoms and Solids for Protons},
At. Data Nucl. Data Tables \textbf{53}, 271 (1993).

\bibitem{Cab08} R. Cabrera Trujillo, J.R. Sabin, E. Deumens, and Y.
Ahrn, \textit{Cross sections for H + and H atoms colliding with Li
in the low-keV-energy region}, Phys. Rev. A \textbf{78}, 012707
(2008).

\bibitem{Ep95} Ch. Eppacher, R. Diez Muino, D. Semrad, and A. Arnau,
\textit{Stopping power of lithium for hydrogen projectiles},  Nucl.
Instrum. Methods Phys. Res.  B \textbf{96}, 639 (1995).


\end{thebibliography}
\end{document}